# Generation of non-classical optical fields by a beam splitter with second-order nonlinearity

*Hari Prakash[1, \*] and Devendra Kumar Mishra[1, 2, \*\*]*

[1]Physics Department, University of Allahabad, Allahabad 211002 (India).

[2]Physics Department, V. S. Mehta College of Science, Bharwari, Kaushambi 212201, U P (India).

E-mails: *prakash_hari123@rediffmail.com; **kndmishra@rediffmail.com

**Abstract**

We propose quantum-mechanical model of a beam splitter with second-order nonlinearity and show that non-classical features such as squeezing and sub-Poissonian photon statistics of optical fields can be generated in output fundamental and second harmonic modes when we mix coherent light beams via such a nonlinear beam splitter.

# 1. Introduction

Non-classical features of optical field such as squeezing, antibunching, and sub-Poissonian photon statistics, have been intensively investigated in quantum optics [1]. Many of these features are now well understood theoretically and observed experimentally and attract considerable attention not only because they cast a new light on some fundamental questions about quantum electrodynamics, but also because of many practical applications such as high-precision measurements [2], optical communications [3], optical information processing [4]. Such applications rely on the possibility of reducing the quantum fluctuations of light below the so-called standard quantum limit. Generation of these non-classical features still remain an open question, notwithstanding a number of attempts have been done since their introduction [1].

Beam splitter, an important device in quantum optical experiments, has been used to change one type of non-classicality to another [5]. It is one of the most widely used innocent looking optical components in any quantum optics laboratory that are understood completely by classical treatment using only linear interactions. Also several attempts have been made in order to understand the quantum mechanical behavior of the lossless beam splitter [6]. Beam splitters are the key elements in interferometers [7], frequently used in the detection of non-classical features, squeezing, and recently it is becoming an important device in the context of rapidly growing quantum information technology as it can generate entanglement [8]. These developments in the study of beam splitters are based on the assumption that it is made up of such a material so that it behave as a linear lossless device. An ideal beam splitter is a *reversible*, *lossless* device in which two incident beams may interfere to produce two emerging beams. Excitations of physical phenomena does not lie in its linearization but in its nonlinearization and, therefore, one may ask a natural question that what will happen when the beam splitter is made up of nonlinear material? How, then, the input / output mode operators will be related? Can a beam splitter, which is made up of nonlinear material, generate non-classical features of optical field in itself if one injects classical light beam through it? These are the basic questions, which prompted us to investigate the quantum mechanical behavior of a beam splitter that is made up of a nonlinear material.

Prior to going into detailed about our work, let us first make a brief look on the developments in the field of nonlinear beam splitter studies. Some attempts have been made to study the problem of nonlinear beam splitter found in the literature. Deutsch [9] gave a very nice scheme of a lossless Kerr nonlinear beam splitter and proposed that a weak Kerr-nonlinear beam splitter might serve to generate non-classical fields. On detailed investigation of this paper, we [10] found that the transformed annihilation and creation operators do not obey the bosonic commutation relation even up to first order in the nonlinearity, although Deutsch points out that each transformed operator must obey separately the bosonic commutation relation. Belinsky & Granovskiy [11] studied about separation of quantum fluctuations into amplitude and phase fluctuations and predicted possible applications in rapidly developing quantum cryptography. Belinsky & Shulman [12] presented a review on beam splitter exhibiting Kerr nonlinearity. Perinova, Luks and Krepelka [13] introduced beam splitter with second-order nonlinearity with the aid of differential equations for co-propagation and for counter-propagation and showed its use for the sub-Poissonian statistics. Pezzé et al. [14] studied a Bose-Einstein condensate beam splitter, realized with a double well potential of tunable height.

Intent of the present letter is to report a scheme of 'beam splitter having second order nonlinearity' [15] upto the lowest order in perturbation, we abbreviate it as the BSSN throughout the letter, and assume that the BSSN is made up of noncentrosymmetric crystals which lack inversion symmetry and, therefore, possess a non-zero $\chi^{(2)}_{ijk}$ [16]. It is found that squeezing and sub-Poissonian photon statistics of optical fields are generated in output second-harmonic mode when we mix coherent light beams via such a nonlinear beam splitter. This result is remarkable because of our simplicity of the model of BSSN.

## 2. The Beam Splitter Model with Second-Order Nonlinearity (BSSN)

A lossless beam splitter is a linear four-port device in which the two radiation modes enter, interfere with each other, and leave it. If $\hat{a}$ and $\hat{b}$ are the boson annihilation operators of the input modes and that of output modes are $\hat{c}$ and $\hat{d}$, the input/output relation [17] of the linear beam splitter, in a more simpler way, is

$$\hat{c} = t\hat{a} + ir\hat{b}; \quad \hat{d} = t\hat{b} + ir\hat{a} \tag{1}$$

where $r$ and $t$ are the real amplitude-reflection and transmission coefficients, obeying $r^2 + t^2 = 1$, of the beam splitter. Such a mixing at the linear beam splitter is described by a unitary transformation [18].

For linear materials, relation between macroscopic polarization $\vec{P}(\vec{r},t)$ and electric field $\vec{E}$ is : $\vec{P}(\vec{r},t) = \varepsilon_0 \chi \vec{E}$, where $\chi$ is the susceptibility tensor characterizing the material. Nonlinear response of the medium to an exciting electric field may be thought of as originating from a nonharmonic potential of the electrons in the media. As such a nonlinear response is usually small effect, it can be captured well by a Taylor expansion of the Polarization $\vec{P}(\vec{r},t)$ induced in the medium in terms of the electric field $\vec{E}$, in the form [19]

$$P_i(\omega_1) = \chi^{(1)}_{ij}(\omega_1, -\omega_1)E_j(\omega_1) + \chi^{(2)}_{ijk}(\omega_1, \omega_1 - \omega_2, \omega_2)E_j(\omega_1 - \omega_2)E_k(\omega_2)$$

$$+ \chi_{ijkl}^{(3)}(\omega_1, \omega_1 - \omega_2 - \omega_3, \omega_2, \omega_3) E_j(\omega_1 - \omega_2 - \omega_3) E_k(\omega_2) E_l(\omega_3) \quad (2)$$

The lowest-order nonlinear contribution to energy is represented by $\chi_{ijk}^{(2)}$. Noncentrosymmetric crystals can possess a nonvanishing $\chi_{ijk}^{(2)}$ tensor which is responsible for the generation of light at the second-harmonic frequency. Second-harmonic generation, also known as frequency doubling, is a process in which a mode at frequency $\omega$ (with annihilation operator $\hat{a}$) couples through a nonlinearity with a mode at frequency $2\omega$ (with annihilation operator $\hat{A}$) and it was first achieved in laboratory in 1961 [20]. The Hamiltonian describing this process for a lossless medium is ($c = \hbar = 1$), $\hat{H} = \omega \hat{a}^\dagger \hat{a} + 2\omega \hat{A}^\dagger \hat{A} + g(\hat{a}^{\dagger 2} A + \hat{a}^2 \hat{A}^\dagger)$, where g is the coupling constant between the two modes which contains the nonlinear susceptibility $\chi^{(2)}$ [21]. Then $[\hat{N}_a + 2\hat{N}_A, \hat{H}] = 0$ and hence the sum $\hat{N}_a + 2\hat{N}_A$ is a constant of motion, $\hat{N}_a = \hat{a}^\dagger \hat{a}$, and $\hat{N}_A = \hat{A}^\dagger \hat{A}$. Therefore, two photons of the fundamental mode are absorbed for every harmonic photon emitted. We denote annihilation operators of fundamental modes by small letters and second harmonic mode by capital letters. It is assumed that the phase matching requirement hold and coupling constant g is appreciable.

If the beam splitter is made up of a material such that it mixes fundamental mode $\hat{a}$ as well as its second harmonic mode $\hat{A}$ with that of mode $\hat{b}$ and its second harmonic mode $\hat{B}$ (Fig. 1), then the output modes with corresponding operators $(\hat{c}, \hat{C})$ and $(\hat{d}, \hat{D})$ can be written as,

$$\hat{c} = t_f \hat{a} + i r_f \hat{b} + z_1 \hat{a}^\dagger \hat{A} + z_2 \hat{a}^\dagger \hat{B} + z_3 \hat{b}^\dagger \hat{A} + z_4 \hat{b}^\dagger \hat{B} \quad (3a)$$

$$\hat{d} = t_f \hat{b} + i r_f \hat{a} + z_4 \hat{a}^\dagger \hat{A} + z_3 \hat{a}^\dagger \hat{B} + z_2 \hat{b}^\dagger \hat{A} + z_1 \hat{b}^\dagger \hat{B} \quad (3b)$$

$$\hat{C} = t_s \hat{A} + i r_s \hat{B} + w_1(\hat{a}^2 + \hat{b}^2) + w_2(\hat{a}^2 - \hat{b}^2) + w_3 \hat{a}\hat{b} \quad (3c)$$

$$\hat{D} = t_s \hat{B} + i r_s \hat{A} + w_1(\hat{a}^2 + \hat{b}^2) - w_2(\hat{a}^2 - \hat{b}^2) + w_3 \hat{a}\hat{b} . \quad (3d)$$

Here $(t_f, t_s)$ and $(r_f, r_s)$ are real transmission and reflection coefficients for amplitudes. Suffixes f and s refer to fundamental and second harmonic modes. These equations involve besides known coefficients $r_{f,s}$ and $t_{f,s}$, fourteen unknown real constants in seven coupling coefficients $z_{1,2,3,4}$ and $w_{1,2,3}$. Such a perturbation expansion can be easily justified on the basis of the fact that if the fundamental modes vary according as $e^{-i\omega t}$, then the second harmonic modes will vary as $e^{-2i\omega t}$. To find the complex coupling constants z and w, we make use of the following properties:

(i) Commutation relations of boson operators to first order in z and w, viz,

$[\hat{c},\hat{c}^\dagger]=1$, $[\hat{d},\hat{d}^\dagger]=1$, $[\hat{c},\hat{d}]=0$, $[\hat{c},\hat{d}^\dagger]=0$, $[\hat{C},\hat{C}^\dagger]=1$, $[\hat{D},\hat{D}^\dagger]=1$, $[\hat{C},\hat{D}]=0$,

$[\hat{C},\hat{D}^\dagger]=0$, $[\hat{c},\hat{D}]=0$, $[\hat{d},\hat{D}]=0$, $[\hat{C},\hat{d}]=0$, $[\hat{D},\hat{d}]=0$.

(ii) Energy conservation, as expressed by

$$\hat{c}^\dagger\hat{c}+\hat{d}^\dagger\hat{d}+2(\hat{C}^\dagger\hat{C}+\hat{D}^\dagger\hat{D})=\hat{a}^\dagger\hat{a}+\hat{b}^\dagger\hat{b}+2(\hat{A}^\dagger\hat{A}+\hat{B}^\dagger\hat{B}),$$ and

(iii) Principle of optical reversibility [18, See also 22], which states $\begin{bmatrix}\hat{c}\\\hat{d}\end{bmatrix}=\hat{U}\begin{bmatrix}\hat{a}\\\hat{b}\end{bmatrix}$ giving $\begin{bmatrix}\hat{a}\\\hat{b}\end{bmatrix}=\hat{U}^\dagger\begin{bmatrix}\hat{c}\\\hat{d}\end{bmatrix}$ for the linear beam splitter case, $\hat{U}$ being the unitary matrix, necessitating the change $\hat{a}\leftrightarrow\hat{c}$, $\hat{b}\leftrightarrow\hat{d}$, $\hat{A}\leftrightarrow\hat{C}$, $\hat{B}\leftrightarrow\hat{D}$, $i\to -i$ *which results in* $z_j\leftrightarrow z_j^*$ and $w_j\leftrightarrow w_j^*$.

For simplicity of calculations, let us take $t_f=t_f=1/\sqrt{2}$, but keep the option of different transmission/reflection coefficients for the second harmonic and write $t_s=\cos\vartheta$, $r_s=\sin\vartheta$. Then, Eqs. (3) can be written as

$$\hat{c}=\tfrac{1}{\sqrt{2}}(\hat{a}+i\hat{b})+z_1\hat{a}^\dagger\hat{A}+z_2\hat{a}^\dagger\hat{B}+z_3\hat{b}^\dagger\hat{A}+z_4\hat{b}^\dagger\hat{B} \tag{4a}$$

$$\hat{d}=\tfrac{1}{\sqrt{2}}(\hat{b}+i\hat{a})+z_1\hat{b}^\dagger\hat{B}+z_2\hat{b}^\dagger\hat{A}+z_3\hat{a}^\dagger\hat{B}+z_4\hat{a}^\dagger\hat{A} \tag{4b}$$

$$\hat{C}=\hat{A}\cos\vartheta+i\hat{B}\sin\vartheta+w_1(\hat{a}^2+\hat{b}^2)+w_2(\hat{a}^2-\hat{b}^2)+w_3\hat{a}\hat{b} \tag{4c}$$

$$\hat{D}=\hat{B}\cos\vartheta+i\hat{A}\sin\vartheta+w_1(\hat{a}^2+\hat{b}^2)-w_2(\hat{a}^2-\hat{b}^2)+w_3\hat{a}\hat{b}. \tag{4d}$$

Using Eqs (4), we get

$$\hat{a}=\tfrac{1}{\sqrt{2}}(\hat{c}-i\hat{d})-\tfrac{1}{2}[\hat{c}^\dagger\hat{C}\,(z_1e^{-i\vartheta}+z_2e^{-i\vartheta}+iz_3e^{i\vartheta}-iz_4e^{i\vartheta})+\hat{c}^\dagger\hat{D}\,(z_1e^{-i o}+z_2e^{-i\vartheta}-iz_3e^{i\vartheta}+iz_4e^{i\vartheta})$$
$$+\hat{d}^\dagger\hat{C}\,(iz_1e^{i\vartheta}-iz_2e^{i\vartheta}+z_3e^{-i\vartheta}+z_4e^{-i\vartheta})+\hat{d}^\dagger\hat{D}\,(-iz_1e^{i\vartheta}+iz_2e^{i\vartheta}+z_3e^{-i\vartheta}+z_4e^{-i\vartheta}) \tag{5}$$

The principle of optical reversibility [18, See also 22] allows us to equate Eq. (5) with

$$\hat{a}=\tfrac{1}{\sqrt{2}}(\hat{c}-i\hat{d})+z_1^*\hat{c}^\dagger\hat{C}+z_2^*\hat{c}^\dagger\hat{D}+z_3^*\hat{d}^\dagger\hat{C}+z_4^*\hat{d}^\dagger\hat{D} \tag{6}$$

which is obtained from Eq. (4a) by making transformations $\hat{a}\leftrightarrow\hat{c}$, $\hat{b}\leftrightarrow\hat{d}$, $\hat{A}\leftrightarrow\hat{C}$, $\hat{B}\leftrightarrow\hat{D}$, $i\to -i$ *which results in* $z_j\leftrightarrow z_j^*$ and $w_j\leftrightarrow w_j^*$. If we define $Z_j=X_j+iY_j\equiv z_j\exp\{i(\pi-\vartheta)/2\}$, the comparison of Eq. (5) and Eq. (6) gives $Y_1=-Y_2$, $Y_3=-Y_4$, and

$$Z_1^*-Z_2^*-Z_3^*+Z_4^*=(Z_1-Z_2-Z_3+Z_4)\exp\{i(4\vartheta-\pi)/2\}. \tag{7}$$

If we put $Z_1 - Z_2 - Z_3 + Z_4 = 4(X_0 + iY_0) = 4R_0 e^{i\varphi}$, defining real constants $\varphi = \frac{\pi}{4} - \vartheta$, $R_0$, $X_0$ and $Y_0$ real, Eq. (7) gives, $Y_1 - Y_3 = 2Y_0$ and $X_1 - X_2 - X_3 + X_4 = 4X_0$. Similarly, if we put $Y_1 + Y_3 \equiv 2Y$, and define a real constant Y, we have

$$Y_1 = Y + R_0 \sin\varphi = -Y_2, \quad Y_3 = Y - R_0 \sin\varphi = -Y_4, \text{ and } X_1 - X_2 - X_3 + X_4 = 4R_0 \cos\varphi. \tag{8}$$

The commutation relation $[\hat{c},\hat{d}] = 0$ provides us one more result $Z_1 - Z_2 - iZ_3 + iZ_4 = 0$, i.e., $X_3 - X_4 = 2Y + 2R_0 \sin\varphi = Y_1 - Y_2$, and $X_1 - X_2 = -2Y + 2R_0 \sin\varphi = Y_4 - Y_3$ and hence $X_1 - X_2 - X_3 + X_4 = -4Y = 4X_0 = 4R_0 \cos\varphi$, i.e., $Y = -R_0 \cos\varphi$. These gives

$$Y_1 = R_0(\sin\varphi - \cos\varphi) = -\sqrt{2}\, R_0 \sin\vartheta, \tag{9a}$$

$$Y_2 = -Y_1 = \sqrt{2}\, R_0 \sin\vartheta, \tag{9b}$$

$$Y_3 = -\sqrt{2}\, R_0 \cos\vartheta, \tag{9c}$$

$$Y_4 = -Y_3 = \sqrt{2}\, R_0 \cos\vartheta, \tag{9d}$$

$$X_1 - X_2 = 2\sqrt{2}\, R_0 \cos\vartheta, \tag{9e}$$

and

$$X_3 - X_4 = -2\sqrt{2}\, R_0 \sin\vartheta, \tag{9f}$$

and define real coefficients $X_{12}$ and $X_{34}$ given by

$$X_{1,2} = X_{12} \pm \sqrt{2}\, R_0 \cos\vartheta \ \& \ X_{3,4} = X_{34} \mp \sqrt{2}\, R_0 \sin\vartheta. \tag{10}$$

All these enable us to simplify and write

$$z_{1,2} = X_{12} e^{-i(\pi-\vartheta)/2} \pm \sqrt{2}\, R_0 e^{-i(\pi+\vartheta)/2} \ \& \ z_{3,4} = X_{34} e^{-i(\pi-\vartheta)/2} \mp i\sqrt{2}\, R_0 e^{-i(\pi+\vartheta)/2}. \tag{11}$$

Application of principle of optical reversibility and commutation relations, then, reduce the number of eight unknown real constants represented by four $z_i$'s to three, viz., $X_{12}$, $X_{34}$, and $R_0$.

Now, applying the same procedure as above, from Eq. (4c) and Eq. (4d) we get for the case second-harmonic mode $\hat{A} = \hat{C}\cos\vartheta - i\hat{D}\sin\vartheta + \frac{i}{2}w_3(\hat{c}^2 + \hat{d}^2)e^{-i\vartheta} - w_2(\hat{c}^2 - \hat{d}^2)e^{i\vartheta} + 2iw_1\hat{c}\hat{d}e^{-i\vartheta}$ and similar application of principle of optical reversibility [18, See also 22], we get $w_1^* = \frac{i}{2}w_3 e^{-i\vartheta}$, $w_2^* = -w_2 e^{i\vartheta}$, $w_3^* = 2iw_1 e^{-i\vartheta}$. Taking $w_2 = iMe^{i\vartheta/2}$, M being a real constant, we can write

$$\hat{C} = \hat{A}\cos\vartheta + i\hat{B}\sin\vartheta + w_1(\hat{a}^2 + \hat{b}^2) + iM(\hat{a}^2 - \hat{b}^2)e^{i\vartheta/2} - 2iw_1^* \hat{a}\hat{b}e^{i\vartheta}, \tag{12a}$$

$$\hat{D} = \hat{B}\cos\vartheta + i\hat{A}\sin\vartheta + w_1(\hat{a}^2 + \hat{b}^2) - iM(\hat{a}^2 - \hat{b}^2)e^{i\vartheta/2} - 2iw_1^* \hat{a}\hat{b}e^{i\vartheta}. \tag{12b}$$

But for our BSSN model, the restriction of energy conservation given by Eq. (4) gives us $M = R_0 = 0$ and $w_1 = \frac{1}{2\sqrt{2}}(X_{34} - iX_{12})\exp[i\vartheta/2]$, and, therefore, $z_1 = z_2 = X_{12}\exp[-i(\pi - \vartheta)/2]$ & $z_3 = z_4 = X_{34}\exp[-i(\pi - \vartheta)/2]$. These results can be used to write, with $X_{34} - iX_{12} \equiv 2\sqrt{2}\kappa\exp(i\eta)$, the input/output relations for BSSN as

$$\hat{c} = \tfrac{1}{\sqrt{2}}(\hat{a} + i\hat{b}) - 2i\sqrt{2}\kappa(-\hat{a}^\dagger \sin\eta + \hat{b}^\dagger \cos\eta)(\hat{A} + \hat{B})e^{i\vartheta/2}, \tag{13a}$$

$$\hat{d} = \tfrac{1}{\sqrt{2}}(\hat{b} + i\hat{a}) - 2i\sqrt{2}\kappa(-\hat{b}^\dagger \sin\eta + \hat{a}^\dagger \cos\eta)(\hat{A} + \hat{B})e^{i\vartheta/2}, \tag{13b}$$

$$\hat{C} = \hat{A}\cos\vartheta + i\hat{B}\sin\vartheta + \kappa(\hat{a}^2 + \hat{b}^2)e^{i(\vartheta+2\eta)/2} - 2i\kappa\hat{a}\hat{b}e^{i(\vartheta-2\eta)/2}, \tag{13c}$$

$$\hat{D} = \hat{B}\cos\vartheta + i\hat{A}\sin\vartheta + \kappa(\hat{a}^2 + \hat{b}^2)e^{i(\vartheta+2\eta)/2} - 2i\kappa\hat{a}\hat{b}e^{i(\vartheta-2\eta)/2}. \tag{13d}$$

Thus, finally, we are left with only two constants $\kappa$ and $\eta$ which depend on the nonlinearity material of which the BSSN is made up of. Eqs. (13) are the input/output relations of the BSSN having second order nonlinearity if we mix two two-mode light beams (one is in fundamental whereas another is in second harmonic mode).

## 3. Generation of non-classical features by BSSN and discussion of the results

For radiation with annihilation operator $\hat{a}$ and with general quadrature $\hat{X}_\theta = \tfrac{1}{2}(\hat{a}^\dagger e^{i\theta} + \hat{a}e^{-i\theta})$, the radiation will exhibit squeezing if $\langle(\Delta\hat{X}_\theta)^2\rangle - \tfrac{1}{4} < 0$ and sub-Poissonian photon statistics characterized by a negative value of Mandel's Q-parameter, $Q \equiv (\langle(\Delta\hat{N})^2\rangle - \langle\hat{N}\rangle)/\langle\hat{N}\rangle$ with $\hat{N} = \hat{a}^\dagger\hat{a}$.

Now we investigate the possibilities of generating these non-classical features by BSSN in the output fundamental as well as the output second-harmonic mode at port 'c', if we mix two coherent beams in states $|\alpha\rangle$, $|\beta\rangle$ in the fundamental modes and the input second-harmonic modes are in vacuum state, i.e., no second harmonic signal is present at both of the inputs. For simplicity of calculations, we take real amplitudes of coherent states, $\alpha \equiv x$ and $\beta \equiv y$, x and y being real positive quantities. Then, the squeezing in output fundamental mode (say 'c') occur if

$$\langle(\Delta\hat{X}_\theta)^2\rangle - \tfrac{1}{4} = \kappa(x + y)[\cos(\eta + \tfrac{1}{2}\vartheta - \theta) + 2\kappa(x + y)] < 0, \tag{14}$$

i.e., if $2\kappa(x + y) < 1$ for $\theta = \eta + \tfrac{1}{2}\vartheta$. Sub-Poissonian photon statistics in output fundamental mode (say 'c') occur if

$$Q = \tfrac{1}{2}[1 - 16\kappa^2(x + y)^2] < 0, \tag{15}$$

i.e., if $4\kappa(x+y) > 1$. Sub-Poissonian photon statistics in output second harmonic mode (say for 'C') occur for negative value of

$$Q = 16\kappa^2(x^2+y^2+4xy\sin 2\eta)(x^2+y^2)x^2y^2/[x^4+y^4+6x^2y^2 \quad 4xy(x^2+y^2)\sin 2\eta], \tag{16}$$

i.e., if $x^2+y^2+4xy\sin 2\eta < 0$. A simple example may be x = y and $-1 < \sin 2\eta < -\frac{1}{2}$, i.e., $-\frac{\pi}{4} < \eta < -\frac{\pi}{12}$. Squeezing in output second harmonic mode (say for 'C') occur for negative value of

$$\left\langle (\Delta \hat{X}_\theta)^2 \right\rangle - \tfrac{1}{4} = \tfrac{1}{2}\kappa^2(x^2+y^2-4xy)(x^2+y^2)(1+\sin 2\eta) \tag{17}$$

Therefore, squeezing can occur in output second-harmonic mode if $(x^2+y^2-4xy) < 0$ and we take $\theta = \tfrac{1}{2}\vartheta - \tfrac{1}{4}\pi$, *irrespective of the BSSN property which is determined by* $\eta$ *provided* $\sin 2\eta \neq -1$.

Therefore, we can see that the non-classical features such as squeezing and sub-Poissonian photon statistics can be generated if we mix two coherent beams via the nonlinear beam splitter having second order nonlinearity. Coherent beams with real amplitudes x and y of the fundamental mode and also the corresponding second-harmonic mode in vacuum state, when mixed via the nonlinear beam splitter having second-order nonlinearity, the output fundamental mode (say for 'c') exhibits squeezing if $2\kappa(x+y) < 1$ and it exhibits sub-Poissonian photon statistics if $4\kappa(x+y) > 1$. We note that squeezing is seen to occur at low intensities and sub-Poissonian statistics at larger intensities.

Further, the output second-harmonic mode (say for 'C') exhibits sub-Poissonian photon statistics if $x^2+y^2+4xy\sin\eta < 0$, which may occur for x = y and $-1 < \sin 2\eta < -\tfrac{1}{2}$. It exhibit squeezing if $(x^2+y^2-4xy) < 0$. This should be important because second-harmonic generation is relatively straightforward compared to other nonlinear interactions and it can have relatively high conversion efficiency. The conversion efficiency is defined as the ratio of second-harmonic power generated to fundamental power input. In order to make such a beam splitter, and the study of this letter relevant, all that we need is that the modes have polarizations in one direction only. Thus, if we have all input modes (two fundamental and two second harmonic) polarized along, say, z-direction, we need generation of nonlinear polarization also only along z-direction. Hence, we must consider materials for BSSN which have $\chi_{zzz} \neq 0$ but $\chi_{xzz}$ and $\chi_{yzz} = 0$. Many such examples of non-centrosymmetric crystals exist [15]. One such example is an orthorhombic mm2 crystal. Contributions from these materials to second harmonic generation can be due to bulk dipole and quadrupole terms, and/or surface dipole terms resulting from broken inversion symmetry [23]. All gases, amorphous materials like glass or polymers, and a large number of crystalline materials do not exhibit this type of nonlinear response to an external electrical field. Second harmonic generation is a phenomenon for optical frequency conversion in which the color of light is changed has numerous applications in physics and technology. Ability to manipulate the color of light, and in particular, quantum states of light, can be an extremely important resource from the perspective of quantum information processing [24]. Masada [25] introduced the development of frequency doubler as a pump source for squeezer for applications in quantum radar. It is expected that according to this study, therefore, BSSN may play an important role in quantum optical experiments which are becoming of intense interest in the study of optical quantum information processing.

**Acknowledgement:** Authors acknowledge fruitful discussions and suggestions with Prof. N. Chandra and Prof. R. Prakash.

**Caption for figure:**

Fig. 1. Scheme of mode transformations via beam splitter with second-order nonlinearity (BSSN).

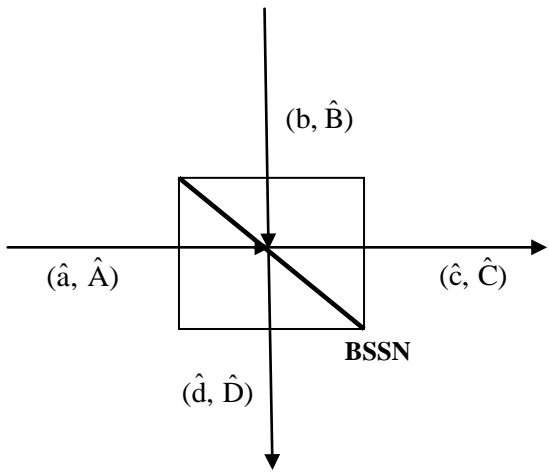

**Fig. 1**